\documentclass[submission,copyright,creativecommons]{eptcs}
\providecommand{\event}{MBT'2014} 

\usepackage{alltt}

\usepackage{epsfig}

\usepackage{floatflt}

\usepackage{array}
\usepackage{xspace}

\usepackage{multicol}

\usepackage[english]{babel}

\usepackage{color}
\usepackage{multirow}

\usepackage[cmex10]{amsmath}
\usepackage{stmaryrd}

\usepackage{latexsym}

\usepackage{listings}

\usepackage{url}

\newtheorem{example}{Example}
\newtheorem{definition}{Definition}

\title{Coverage Criteria for Model-Based Testing\\using Property Patterns}
\author{Kalou Cabrera Castillos
\institute{LAAS-CNRS\\BP 54200\\31031 Toulouse cedex 4, France}
\email{kcabrera@laas.fr}
\and
Fr\'ed\'eric Dadeau \qquad\qquad Jacques Julliand
\institute{FEMTO-ST Institute/INRIA CASSIS Project \\ 16 route de Gray \\25030 Besan\c{c}on cedex, France}
\email{\quad frederic.dadeau@femto-st.fr \quad\qquad jacques.julliand@femto-st.fr}
}
\def\titlerunning{Coverage Criteria for MBT using Property Patterns}
\def\authorrunning{K. Cabrera Castillos, F. Dadeau \& J. Julliand}
\begin{document}
\maketitle

\begin{abstract}
We present in this paper a model-based testing approach aiming at generating test cases from a UML/OCL model and a given test property. The property is expressed using a dedicated formalism based on patterns, and automatically translated into an automaton. 
We propose new automata coverage criteria that are tailored to the property automata we consider. These criteria are based on the coverage of a relevant subset of the transitions related to the original property, aiming at producing test cases that illustrate the dynamics of the system described in the property. In addition, we propose to produce test cases that can ensure the robustness of the system w.r.t. the property, by mutating the property automaton, in order to try to provoke events that would be forbidden by the property. This approach has been implemented into a large tool set and it has been experimented on realistic case studies, in the context of industrial research projects.



\end{abstract}

\section{Introduction and Motivations}
\label{sec:intro}
Model-Based Testing is one of the best ways to automate software testing. 
Even if the design of the model is a costly task, it is damped by the profits 
that can be made from it. Indeed, the model can be used to automatically 
compute the test cases, from simple input test data to complete sequences 
involving several steps. In addition, it is used to provide the oracle,
namely the expected results of the test, thus deciding of the conformance 
between the system under test and its formal representation. 
Many model-based software testing tools exist, such as 
Smartesting's CertifyIt~\cite{certifyit}, Conformiq Designer~\cite{conformiq}, 
or STG~\cite{stg}. 

In practice, some of these tools work by automatically applying a structural
test selection criterion on the model (such as the structural coverage of the OCL 
code~\cite{OCL} of the operations contained in an UML class diagram). Even 
if this approach is quite effective in practice, it suffers from its subjectivity and 
thus, specific behaviours of the system, which require a more intensive test effort, 
are not much targeted by this testing strategy. To overcome this problem, dynamic 
test selection criteria are introduced. They consist in scenario-based testing
approaches that aim at exercising more extensively specific parts of the
considered system. 

In general, test scenarios are expressed in a dedicated language that can be
either textual~\cite{ccdj11:ij} or graphical~\cite{stg}, describing sequences 
of steps (usually operation calls) that can be performed, along with possible 
intermediate states reached during the unfolding of the scenario. 
Nevertheless, the design of the test scenarios remains a manual task that 
we aim to automate. 
During previous experiments in the use of scenarios, we have noticed that 
scenarios often originate from a manual interpretation of a given property
that exercises the dynamics of the system~\cite{POSE}. 
Our goal is now to express such properties, in a simple formalism, that can 
be later exploited for testing purposes. 
To achieve that, Dwyer \emph{et al.} introduced the notion of \emph{property 
patterns} that makes it possible to express dynamic behaviours of the systems
without employing complex temporal logics formalism~\cite{DAC99}. 
Such properties are
expressed with a \emph{scope}, that delimits some fragments of the execution
of the system, and a \emph{pattern} that expresses occurrences, absences or
chains of given events inside the considered scope. 

\begin{figure}[tb]
\centering
\includegraphics[width=12cm]{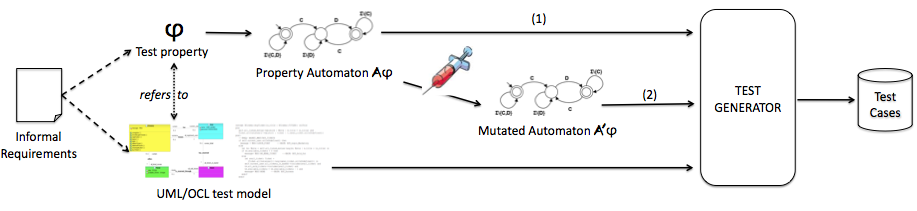}
\vspace{-0.3cm}
\caption{\label{fig:process}Process of the property-based testing approach}
\vspace{-0.4cm}
\end{figure}

The proposed process is depicted in Fig.~\ref{fig:process}. A test property $\varphi$ is described 
using a property pattern that holds on the UML/OCL model. This property is translated into an
automaton $\mathcal{A}_\varphi$, which captures the events that are described
in the test property. Such an automaton represents a monitor of the satisfaction of the property 
during the execution of the system/model. It can be used to 
measure the coverage of the property by existing test suites, as explained
in~\cite{ICTSS11}, but also to produce test 
cases that aim at exercising the property according to dedicated coverage criteria. 
We introduce, in a first contribution, dedicated property automata coverage  
criteria, inspired from classical automata coverage criteria, that aim at characterizing 
relevant tests 
highlighting the behaviours described in the property. In this context, the tests are
directly extracted from the automaton as is (arrow $1$), to cover the events of the property.
As a second contribution, we propose to refine the property automaton so as to exhibit 
possible violations of the property, and derive test cases that aim at ensuring the robustness 
of the system w.r.t. the property. To this end, we apply dedicated mutation operators (arrow $2$) 
on the transitions of the original automaton $\mathcal{A}_\varphi$, producing a new 
automaton~$\mathcal{A'}_\varphi$ that is then exploited similarly. 

This paper is organized as follows. Section~\ref{sec:umlocl} presents the 
considered UML subset that is used for our model-based testing approach and
introduces a running example. Then, 
Section~\ref{sec:propertyautomata} presents the property formalization language 
and its associated semantics as automata. The first contribution, namely the property 
automata 
coverage criteria, is presented in Section~\ref{sec:automatacoverage}. The second
contribution, namely the mutation of automata transitions, is explained in 
Section~\ref{sec:mutation}. The experimental assessment of this approach is 
reported in Section~\ref{sec:experiments}. Then, Section~\ref{sec:relatedworks} 
compares our approach with related works. Finally, Section~\ref{sec:conclusion} 
concludes and presents the future works.

\section{UML/OCL Models and Running Example}
\label{sec:umlocl}

This section presents the UML/OCL models that are considered in our approach, along with 
a running example that will be used to illustrate the contributions in the rest of the paper.  
As our approach is technically bound to the CertifyIt test generation engine, we focus on the 
subset of UML considered by this tool. However, the approach could be applicable to any other
modelling language. 
 
\subsection{UML4ST -- a subset of UML for Model-Based Testing}
\label{subsec:uml4st}

The UML models we consider are those supported by the CertifyIt test generator, 
commercialized by the Smartesting company.
This tool automatically produces model-based tests from a UML
model~\cite{bglp+07:ip} with OCL code describing the behaviors of the
operations. CertifyIt does not consider the whole UML notation as input, it
relies on a subset named UML4ST (UML for Smartesting) which considers class
diagrams, to represent the data model, augmented with OCL
constraints~\cite{OCL}, to describe the dynamics of the system. It also requires
the initial state of the system to be represented by an object diagram. Finally,
a statechart diagram can be used to complete the description of the system
dynamics.


OCL provides the ability to navigate the model, select collections of objects and 
manipulate them with universal/existential quantifiers to build first-order logic 
expressions. 
Regarding the OCL semantics, UML4ST does not consider the third logical 
value \emph{undefined} that is part of the classical OCL semantics. All 
expressions have to be defined at run time in order to be evaluated. CertifyIt
interprets OCL expressions with a strict semantics, and raises execution errors
when encountering null pointers.
 
These restrictions w.r.t. the classical UML semantics originate from the fact 
that the UML/OCL model aims at being used for Model-Based Testing purposes. 
As such, it requires to use an executable UML/OCL model, since the 
abstract test cases are obtained by animating the model.

\subsection{Running Example}
\label{subsec:example}

 We illustrate the UML/OCL models that are considered using a simple model of a
 web application named eCinema. This application provides a means for registered 
 users to book tickets for movies that are screened in a cinema. 
 
 \begin{figure}[!b]
 \centering
 \includegraphics[width=11cm]{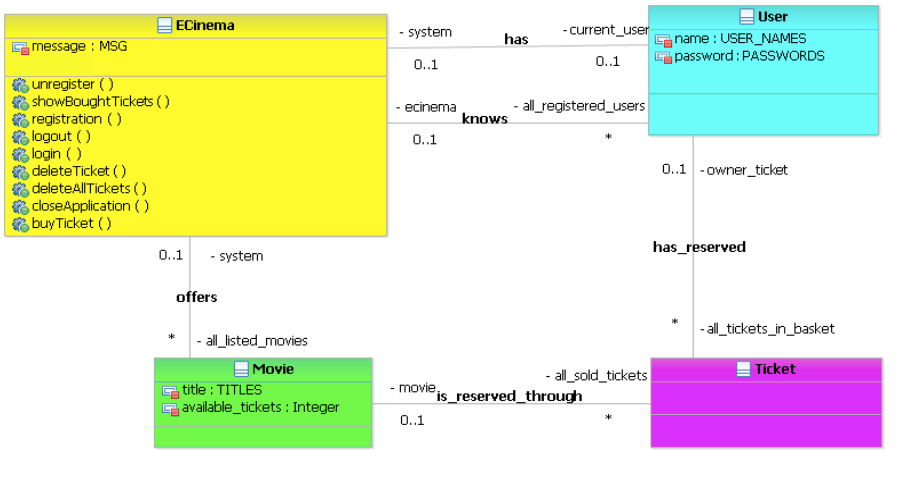}
 \vspace*{-0.4cm} 
 \caption{\label{fig:uml:ecinema:classes}Class diagram of the eCinema model}
 \vspace*{-0.5cm} 
 \end{figure}
 
 The UML class diagram, depicted in Fig.~\ref{fig:uml:ecinema:classes} contains the classes
 of the application: \texttt{ECinema}, \texttt{Movie}, \texttt{Ticket} and \texttt{User}. The 
 \texttt{ECinema} class models the system under test (SUT) and contains the API operations 
 offered by the application. This application proposes classical online booking features: a 
 registered user may login to application, purchase tickets, view his basket, delete one or
 all tickets from his basket, and logout.

 \begin{figure}[!t]
 \begin{center}
 {\scriptsize
 \vspace*{-0.5cm}
 \begin{alltt}
 context ECinema::buyTicket(in_title : ECinema::TITLES): oclVoid
 effect:
 \quad ---@REQ: BASKET_MNGT/BUY_TICKETS
 \quad if self.current_user.oclIsUndefined() then
 \quad\quad message = MSG::LOGIN_FIRST      ---@AIM: BUY_Login_Mandatory
 \quad else
 \quad \quad let tm: Movie = self.all_listed_movies->any(m: Movie | m.title = in_title) in
 \quad \quad if tm.available_tickets = 0 then 
 \quad \quad \quad message= MSG::NO_MORE_TICKET     ---@AIM: BUY_Sold_Out
 \quad \quad else 
 \quad \quad \quad let t: Ticket = (Ticket.allInstances())->any(owner_ticket.oclIsUndefined()) in
 \quad \quad \quad self.current_user.all_tickets_in_basket->includes(t) and
 \quad \quad \quad tm.all_sold_tickets->includes(t) and
 \quad \quad \quad tm.available_tickets = tm.available_tickets - 1 and
 \quad \quad \quad message= MSG::NONE     ---@AIM: BUY_Success
 \quad \quad endif 
 \quad endif
 \end{alltt}}
 \end{center}
 \vspace*{-0.9cm}
 \caption{\label{fig:uml:buyticket}OCL code of the \texttt{buyTicket} operation}
 \end{figure}
 
 Figure~\ref{fig:uml:buyticket} shows the OCL code of the \emph{buyTicket} operation. 
 Upon invocation, the caller provides the title of the movie for which (s)he wants
 to buy a ticket. The operation first checks that the user is registered on the application, and 
 then checks if there exists an unallocated tickets for this movie. If all these
 verifications succeed, a ticket is associated to the user. This operation is specified in a defensive style, and as such,
 its precondition is always true, and the postcondition is in charge of distinguishing nominal 
 cases from erroneous calls. We assume in the rest of the paper that all operations are 
 specified this way, which is realistic since the model is used for testing purposes. 
 
The OCL code of this operation contains non-OCL
annotations, inserted as comments, such as \verb+---@AIM: id+ and \verb+---@REQ: id+. The
\verb+---@AIM:id+ tags denote test targets while the \verb+---@REQ: id+ tags
mark requirements from the informal specifications. These tags can be used to reference
a particular behaviour of the operation (e.g. \texttt{@AIM:BUY\_Login\_Mandatory} represents a 
failed invocation of this operation, due to the absence of a user logged on the system). 
Notice that it is possible to know which tags were covered during the execution of the model, 
inside the test cases, providing a feedback on the structural coverage of the OCL by 
the test cases.

\vspace{-0.3cm}
\section{Property Patterns and Automata}
\label{sec:propertyautomata}
We now present the property language and their associated property automata.

\vspace{-0.3cm}
\subsection{Property Pattern Language}
\label{subsec:property:language}

The property description language is a temporal extension of OCL. This language 
is based on patterns which avoids the use of complex temporal formalisms, such 
as LTL or CTL. 
We ground our work on the initial proposal of Dwyer \emph{et~al.}~\cite{DAC99} in 
which a temporal property is a \textit{temporal pattern}
that holds within a \textit{scope}. Thus, the user can define a temporal property
choosing a pattern and a scope among a list of predefined schema.
The scopes are defined from \textit{events} and delimit the impact of the pattern.
The patterns are defined from events and \textit{state properties} to define 
the execution sequences that are correct. The state properties and the event are described
based on OCL expressions. \\

\noindent \textbf{Patterns.} There are five main temporal patterns:
$(i)$~\texttt{always} $P$ means that state property $P$ is satisfied by any state,
$(ii)$~\texttt{never} $E$ means that event $E$ never occurs,
$(iii)$~\texttt{eventually} $E$ means that event $E$ eventually occurs in a subsequent state,
this pattern can be suffixed by a bound which specifies how many occurrences are expected
(at least $k$, at most $k$, exactly $k$ times),
$(iv)$~$E_1$ (\texttt{directly}) \texttt{precedes} $E_2$ means that event $E_1$ (directly) precedes 
event $E_2$,
$(v)$~$E_1$ (\texttt{directly}) \texttt{follows} $E_2$ means that event $E_2$ is (directly) followed 
by event $E_1$. \\

\noindent \textbf{Scopes.} Five scopes can apply to a temporal pattern $TP$:
$(i)$~$TP$~\texttt{globally} means that $TP$ must be satisfied on any state of the whole execution,
$(ii)$~$TP$~\texttt{before} $E$ means that $TP$ must be satisfied before the first
occurrence of $E$, 
$(iii)$~$TP$~\texttt{after} $E$ means that $TP$ must be satisfied
after the first occurrence of $E$, 
$(iv)$~$TP$~\texttt{between} $E_1$ \texttt{and}
$E_2$ means that $TP$ must be satisfied between any occurrence of $E_1$ followed by an occurrence of $E_2$,
$(v)$~$TP$~\texttt{after} $E_1$ \texttt{until} $E_2$ means that $TP$ must be
satisfied between any occurrence of $E_1$ followed by an occurrence of $E_2$ and after the last occurrence of $E_1$ that 
   is not followed by an occurrence of $E_2$. 

\noindent \textbf{Events.} Scopes and patterns refer to events that can be of two kinds. 
On the one hand, events denoted by
\texttt{isCalled(op, pre, post, \{tags\})} represent operation calls. In this expression, 
\texttt{op} designates the operation name, \texttt{pre} and \texttt{post} are OCL expressions
respectively representing a precondition and a postcondition. Finally, \texttt{\{tags\}} 
represents a set of tags that can be activated by the operation call. 
Such an event is satisfied on a transition when the operation \texttt{op} is 
called from a source state satisfying the precondition \texttt{pre} and leading to a target state
satisfying the postcondition \texttt{post}
and executing a path of the control flow graph of the operation \texttt{op}
which is marked by at least one tag of the set of tags denoted \{\texttt{tags}\}. Notice that 
the three components \texttt{pre}, \texttt{post} and \texttt{\{tags\}} are optional. 
On the other hand, events denoted by \texttt{becomesTrue(}$P$\texttt{)}
where $P$ is a OCL predicate,  
are satisfied by any operation call from a 
state in which $P$~evaluated to false, reaching a state in which $P$~evaluates to true.

\begin{example}[Property Example]\label{ex:property}
Consider the eCinema example given in Sect.~\ref{subsec:example}. An informal access control 
requirement expresses that: ``the user must be logged on the system
to buy tickets''. This can be expressed by the following three properties that put the 
focus on various parts of the execution of the system.

\begin{center}

\begin{tabular}{rlr}
\textbf{never} & isCalled(buyTicket, \{@AIM:BUY\_Success\}) & \multirow{2}{*}{$(Property~1)$}  \\
 \textbf{before} & isCalled(login, \{@AIM:LOG\_Success\}) &  \\
 & &  \\
\textbf{eventually} & isCalled(buyTicket, \{@AIM:BUY\_Success\}) \ \textbf{at least 0 times} & \\
 \textbf{between} & isCalled(login, \{@AIM:LOG\_Success\}) &  $(Property~2)$ \\
 \textbf{and} & isCalled(logout, \{@AIM:LOG\_Logout\}) & \\
 & & \\
\textbf{never} & isCalled(buyTicket, \{@AIM:BUY\_Success\}) & \\
 \textbf{after} & isCalled(logout, \{@AIM:LOG\_Logout\}) & $(Property~3)$ \\
 \textbf{until} & isCalled(login, \{@AIM:LOG\_Success\})  &  \\
\end{tabular} 
\end{center}

First, with Property~$(1)$, we specify that before a first successful login, it is not possible to succeed in 
buying a ticket. Second, we specify that when the user is logged in, he may buy
a ticket (Property~$(2)$).
Notice that this property uses a workaround of the eventually pattern to express an optional action. 
Finally, Property~$(3)$ specifies that it is also impossible to buy a ticket after logging out, unless logging in again.
\end{example}



\subsection{Property Semantics using Automata}
\label{subsec:property:semantics}

The properties are interpreted on executions that are viewed as
sequences of pairs of a state and an event that represent a sequence of transitions.
The semantics of the test properties are expressed by means of automata. Indeed, 
the temporal language is a linear temporal logic whose expression power is included 
in the $\omega$-regular languages. 


The semantics of a temporal property is a labelled automaton
which is defined by Def.~\ref{def:automaton}. The method that associates an
automaton to a temporal property is completely defined in~\cite{cdjkt13:ip}.
This automaton describes the set of accepted 
executions of the property and highlights specific transitions representing 
the events used in the property description. In addition, the automaton may 
contain at most one rejection state that indicates the violation of the
property when reached.

\begin{definition}[Property Automaton]
\label{def:automaton}
Let $\Sigma$ be a set of events.
An automaton is a quintuplet $\mathcal{A}=\langle Q,q_0,F,R,T \rangle$ in which:
 $Q$ is a finite set of  states, $q_0$ is an initial state ($q_0 \in Q$),
 $F$ is a set of final states ($F\subseteq Q$), 
 $R$ is a set of rejection states ($R \subseteq Q$), 
 $T$ is a set of transitions ($T\subseteq Q \times \mathcal{P}(\Sigma)\times Q$) labelled by a set of events.
\end{definition}
 
We call $\alpha-$transitions the transitions of $T$ that are labelled by the events 
expressed in the original property, and we call $\Sigma-$transitions the other transitions.
$\Sigma-$transitions are named after their expression as they are labelled by a restriction
on $\Sigma$ (the set of all possible events). 
 
Notice that, when considering safety properties (something bad never happens), the 
set $R$ of rejection state is necessarily not empty. Notice also that, for a given state 
(resp. transition), it is possible to know if the state (resp. transition) originates from the scope
or the pattern of the property. Notice also that the final states catch that the scope
has been executed at least once. Thus final states are not accepting states as in 
traditional B\"uchi automata; they represent the test goals in the sense that we 
expect test cases to reach such states at some point. 

Events in the automaton are quadruplets $[op,pre,post,\{tags\}]$ in which 
$op$ designates an operation, $pre$ and $post$ respectively denote 
pre- and postconditions, and $tags$ specifies a set of tags. The events
used in the test properties are thus rewritten to match this formalism: 
\texttt{isCalled(}op, pre, post, \{tags\}\texttt{)} rewrites to $[op, pre, post, \{ tags \} ]$
and \texttt{becomesTrue($P$)} rewrites to $[ \_, not(P), P, \_ ]$, in which \_ replaces 
any acceptable value of the corresponding component.


 \begin{figure}[!b]
 \centering
 
\begin{tabular}{ccccc}
   \includegraphics[height=4.3cm]{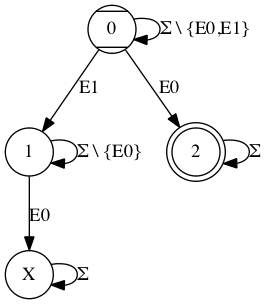}  & ~~~~~~~~ &
   \includegraphics[height=4.3cm]{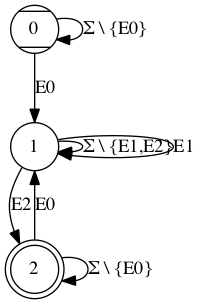} & ~~~~~~ &
   \includegraphics[height=4.3cm]{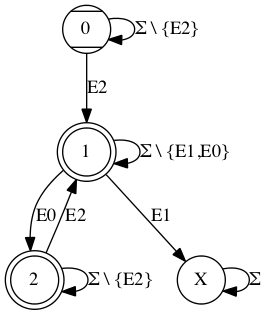} \\
 
 	Property 1 & & Property 2 & & Property 3 \\
 	
 	\\
 	
 	\multicolumn{5}{l}{E0:~[login,\_,\_,\{@AIM:LOG\_Success\}]}\\
 	\multicolumn{5}{l}{E1:~[buyticket,\_,\_,\{@AIM:BUY\_Success\}]} \\
 	\multicolumn{5}{l}{E2:~[logout,\_,\_,\{@AIM:LOG\_Logout\}]}
 	  
 \end{tabular}
  \caption{\label{fig:auto:property}Automata representation for the properties given in Example~\ref{ex:property}}
 \end{figure}

\begin{example}[\label{ex:automaton}Automaton of a Temporal Property] Consider
the property given in Example~\ref{ex:property}. Figure~\ref{fig:auto:property} shows 
the automata representation associated to Property~$1$ (left), Property~$2$ (middle) 
and Property~$3$ (right). 
Notice that the left-hand side (resp. right-hand side) automaton displays an error state, 
identified by ``X'', that can be reached if the system authorizes to perform a ticket 
purchase before logging in (resp. after logging out and before logging in again). The 
automaton of Property~2, in the middle, does not 
display an error state meaning that the property can never be falsified. In addition, it 
exhibits a reflexive transition that represents the optional event, that may or may not 
take place when the user is logged. 

On these automata, the $\alpha-$transitions are
the transitions labelled by events E0, E1 and E2. The other transitions are $\Sigma-$transitions. 
 \end{example}

These automata provide a means for monitoring the satisfaction of the property
by the execution of the model. We assume that the model is correct 
w.r.t. the property. In that sense, only valid traces can be extracted from the model,
and no transition leading to an error state can possibly be activated. 

The next two sections show how to exploit these automata by defining dedicated test coverage 
criteria, that can be used either for evaluating an existing test suite, or for generating new tests
supposed to illustrate or exercise the property.

\vspace{-0.3cm}
\section{Property Automata Coverage Criteria}
\label{sec:automatacoverage}
We present in this section the automata coverage criteria that we propose. These dedicated
coverage criteria 
originate from the observation that classical coverage criteria on automata are not
relevant for our property automata. Indeed, criteria such as transition coverage, or
transition-pair coverage, make no distinction between the transitions of the automata. 
However, in our case, all the transitions of the property automata are not of equal 
importance. For example, consider the automata provided in Fig.~\ref{fig:auto:property}.
Reflexive $\Sigma-$transitions only exist to capture all possible
executions of the model but their sole purpose is to make it possible to put the model
in a state from which the $\alpha-$transitions can be activated. While classical 
coverage criteria would target these $\Sigma$-transitions, we propose new coverage 
criteria focused on $\alpha-$transitions, 
aiming at activating them, but also focusing on different paths which iterate over specific
parts of the automaton. We present these criteria and illustrate, for each of them, their
relevance in terms of property coverage. Before that, 
we start by introducing some preliminary definitions.

\subsection{Preliminary Definitions}

We consider that an abstract test case is defined on the model as a finite sequence of steps,
each of them  
formalized by $s_{i+1}, \vec{o_i}, tags_i \leftarrow op_i(\vec{in_i},s_i)$
(for $i \geq 0$ and $i <$ the length of the test case)
in which $s_i$ (resp. $s_{i+1}$) is the model state before (resp. after) the step, $op_i$ is a
model operation called with inputs $\vec{in}_i$ returning outputs $\vec{o_i}$
and activating the behaviours identified by the $tags_i$ set. We denote by $s_0$ the initial state of the 
model. 

The conversion of a test case (computed from the model) into a path of the automaton 
is made by matching the steps of the test case with the events of the automaton, accordingly 
to the following definition. 

\begin{definition}[Step/Event matching]
A step formalized by $s_{i+1}, \vec{o_i}, tags_i \leftarrow op_i(\vec{in_i},s_i)$  
is said to match an event $[op,pre,post,tags]$ if and only if the four conditions
hold: $(i)$~$op = op_i$ 
or $op$ is undefined (symbol $\_$), $(ii)$~$pre$ is satisfied in $s_i$ (modulo 
substitution of $\vec{in_i}$ in $pre$), $(iii)$~$post$ is satisfied in $s_{i+1}$, and 
$(iv)$~$tags \cap tags_i \neq \emptyset$
\end{definition}

Given a test case, each step $s_{i+1}, \vec{o_i}, tags_i \leftarrow op_i(\vec{in_i},s_i)$ is matched 
against the possible transitions $q \stackrel{e_i}{\rightarrow} q'$ that can be activated
from the current automaton state $q$ (initially, $q_0$ when the first step is considered). When 
a given step/event is matched, the exploration of the automaton restarts
from $q'$ the state targeted by the transition. As the property automata are deterministic and complete, 
there is exactly one transition that can be matched by each step of the test case. 


\subsection{Coverage Criteria for the Property Automata}

We present in this section the four coverage criteria that we propose. Since the model
is expected to satisfy the property, the paths inescapably leading to the error state
are not supposed to be coverable, and thus, their transitions are not considered in the 
coverage criteria that we now present. 



The first two coverage criteria that we propose consider the $\alpha$-transitions. 

\begin{definition}[$\alpha-$transition coverage]
A test suite is said to satisfy the $\alpha-$transition coverage criterion if and only if 
each $\alpha-$transition of the automaton is covered by at least one test case of the 
test suite. 
\end{definition}

This first coverage criterion is an adaptation of the classical transitions-coverage criteria 
from the literature~\cite{Huang75}. It aims at covering the transitions that are labelled by
events initially written in the associated temporal property. A test suite satisfying this 
criterion ensures that all the events expressed at the property level are highlighted by the
test suite. 
 
\begin{example}[$\alpha-$transition coverage]
On the example shown in Fig.~\ref{fig:auto:property}, for Property~2, a test suite 
satisfying the $\alpha-$transition coverage criterion ensures that at least one test case 
illustrates the optional ticket purchase by covering transition $1 \stackrel{E1}{\rightarrow} 1$. 
Also, another test case should illustrate the fact that two iterations of the scope are possible, 
by covering transition $2 \stackrel{E0}{\rightarrow} 1$. 
\end{example}

\begin{definition}[$\alpha-$transition-pair coverage]
A test suite is said to satisfy the $\alpha-$transition-pair coverage criterion if and only if 
each successive pair of $\alpha-$transitions is covered by at least one test case of the test suite. 
\end{definition}

Notice that this criterion considers the coverage of pairs of $\alpha-$transitions 
reaching a particular state, and originating from the same state. However, it is possible to 
display intermediate $\Sigma-$transitions between a pair of $\alpha-$transitions. 

\begin{example}[$\alpha-$transition-pair coverage]
On the example shown in Fig.~\ref{fig:auto:property}, for Property~2, a test suite 
satisfying the $\alpha-$transition coverage criterion ensures the coverage of the following
pairs: $(0 \stackrel{E0}{\rightarrow} 1, 1 \stackrel{E1}{\rightarrow} 1)$, 
$(0 \stackrel{E0}{\rightarrow} 1, 1 \stackrel{E2}{\rightarrow} 2)$, 
$(1 \stackrel{E1}{\rightarrow} 1, 1 \stackrel{E2}{\rightarrow} 2)$, 
$(1 \stackrel{E2}{\rightarrow} 2, 2 \stackrel{E0}{\rightarrow} 1)$, 
$(2 \stackrel{E0}{\rightarrow} 1, 1 \stackrel{E1}{\rightarrow} 1)$ and 
$(2 \stackrel{E0}{\rightarrow} 1, 1 \stackrel{E2}{\rightarrow} 2)$. 
A test suite satisfying this coverage criterion thus ensure the existence of tests illustrating the
buying of a ticket, but also tests performing a login followed by a logout without intermediate
ticket purchase, and also tests illustrating the optional ticket purchase in a second iteration over 
the scope. 
\end{example}

The last two coverage criteria that we propose consider the structure of the property 
and aim at covering internal or external loops inside the property automaton, in order
to iterate over the pattern or the scope of the property. 

\begin{definition}[$k$-pattern coverage]
A test suite is said to satisfy the $k$-pattern-activation coverage criterion if
and only if the $\alpha-$transitions of the pattern of the automaton are iterated 
between 0 and $k$ times,
each loop in the pattern being performed without exiting the pattern-part of the
automaton.  
\end{definition}

This coverage criterion aims at activating the internal loops inside the pattern-part of the
automaton, without covering any transition of scope during these iterations. This coverage
criterion is not applicable to any pattern; it only applies to {\tt precedes}, {\tt follows} and
some forms of the {\tt eventually} pattern. 

\begin{example}[$k$-pattern-coverage]
On the example shown in Fig.~\ref{fig:auto:property}, for Property~2, a test suite 
satisfying the $2$-pattern coverage criterion ensures the coverage of 0, 1, and 2 iterations
of the reflexive $\alpha-$transition $1 \stackrel{E1}{\rightarrow} 1$. 
\end{example}

\begin{definition}[$k$-scope coverage]
A test suite is said to satisfy the $k$-scope-activation coverage criterion if
and only if the $\alpha-$transitions of the scope of the automaton are iterated 
between 1 and $k$ times, 
and covering each time at least one $\alpha-$transition of the pattern. 
\end{definition} 

This coverage criterion aims at activating the external loops outside the pattern-part 
of the automaton. Similarly to the $k$-pattern criterion, the $k$-scope criterion is not 
applicable to all scopes, but restricts its usage to only repeatable ones, 
namely {\tt between} and {\tt after}.

\begin{example}[$k$-scope coverage]
On the example shown in Fig.~\ref{fig:auto:property}, for Property~2, a test suite 
satisfying the $2$-scope coverage criterion ensures the coverage of 1 and 2 logout-login 
sequences, by covering cycle $1 \stackrel{E1}{\rightarrow} 2 \stackrel{E0}{\rightarrow} 1$. 
\end{example}


These four coverage criteria are based on the property automata as is, and thus, will only
illustrate the property and show that they are correctly implemented (e.g. the occurrences 
of events are authorized by the implementation, along with the repetition of scopes, etc.)
However, showing that unexpected events do not appear requires an additional and 
dedicated strategy for robustness testing, that we now present.

\section{Property Automata Mutation for Testing Robustness}
\label{sec:mutation}

The automata coverage criteria described in the previous section focus on
activating events expressed within the test properties. Thus, these coverage
criteria aim at illustrating that the properties are correctly implemented. However, in the
cases of safety properties (something bad should never happen), it might be
interesting to produce test cases that aim at an attempt to violate the
property. The cases in which the property is violated are clearly identified in the automaton, 
being displayed 
through error states. Unfortunately, targeting the activation of transitions leading to 
these error states is irrelevant: since the model (used to compute the tests) is supposed 
to satisfy the property, these transitions can not be activated as is.

In this section, we propose mutation operators that apply to events labelling
the transition leading to error states, so as to make them activable, thus providing
an interesting test target that aims at leading to a possible violation of the
property.
These mutations and the robustness coverage criterion aims to simulate an 
erroneous implementation of the property that would mistakenly allow the 
activation of a unexpected event. As we are performing a Model-Based Testing 
approach, it is thus mandatory to be able to compute a sequence performing 
forbidden events, at the model level.

\subsection{Event Mutation Operators}

Our goal is to provoke unexpected events. As these latter can not be activated on 
the model, the idea is to get closer to the inactivable event. To achieve that, we 
apply mutations on these events. These mutations apply mainly to the uncontrollable
part of the events (postconditions and tags), and keep the controllable part the lesser
modified. 

The mutations we propose modify the transitions of the automata. They target the 
events labelling the transitions, and can be of two kinds: $(i)$~predicate mutation
rules, inspired from classical mutation operators over predicates~\cite{demillo},
applied to pre- and postconditions, and $(ii)$~tag mutation rules applied to the tag 
list of the events. 

\paragraph{Postcondition/Tag Removal.} This rule consists in removing the postcondition
and the tag list from the event.
$$[op,pre,post,T] \leadsto [op,pre,\_,\_]$$
Both tags and postconditions are systematically removed, as these two elements are 
frequently related. Their combined removal thus avoids creating inactivable events.

\paragraph{Precondition Removal.} This rule consists in removing the precondition
of the event. 
$$[op,pre,post,T] \leadsto [op,\_,\_,\_]$$
When applied, this mutation also removes the postcondition and tags, in order to 
weaken the event, and increase the chances that the mutation will produce an 
activable event.  

\paragraph{Predicate Weakening.}
The predicate removal mutation replaces each literal in a conjunction by true. This 
removal applies to both pre- and postconditions.  
$$[op,A \wedge B, C \wedge D, T] \leadsto [op,A,\_,\_], [op,B,\_,\_], [op, A \wedge B, C, \_], [op, A \wedge B, D, \_]$$
When applied to the postcondition, this rule removes the tags from the event. 
If it is applied to the precondition, this rule also removes the postcondition from
the event.


\begin{example}[Event mutation] 
Consider the examples provided on Fig.~\ref{fig:auto:property}, left-hand side or 
right-hand side. In both cases, event $E1=[buyticket,\_,\_,\{@AIM:BUY\_Success\}]$ 
can be rewritten to $E'1=[buyticket,\_,\_,\_]$. This event represents the attempt to perform 
a ticket purchase but without any expectation regarding the success or the failure 
of this operation. 
\end{example}

\subsection{Automata Mutation and Robustness Coverage Criteria}

The mutation operators that we propose can be applied on a given property
automaton~$\mathcal{A}$. The automaton is modified as follows: 
$(i)$~each transition leading to the error state is mutated, and 
$(ii)$~the targeted error state becomes the only final state of the 
new automaton. 
We denote  $\mathcal{A'}$ the new automaton obtained after
mutation. 

\begin{figure}[!b]
\centering
%
\begin{tabular}{ccccccc}
 \multirow{2}{*}{\includegraphics[width=3cm]{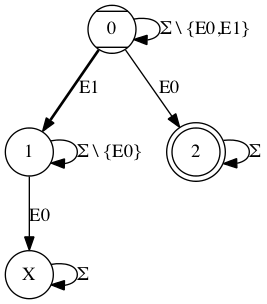}} &
	 &
	\multirow{2}{*}{\includegraphics[width=3cm]{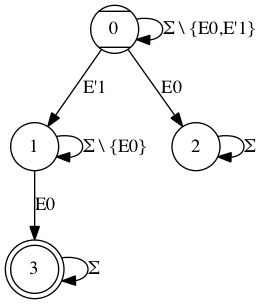}} &
	 ~~~~~ &
	\multirow{2}{*}{\includegraphics[width=3cm]{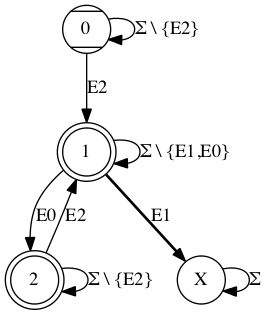}} &
	 &
	\multirow{2}{*}{\includegraphics[width=3cm]{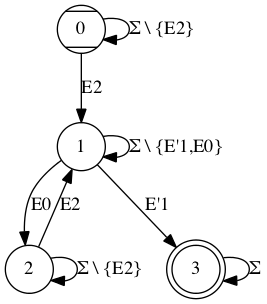}} \\
	\\
	\\
	\\	
	& $\Rightarrow$	& & & & $\Rightarrow$ & \\	
	\\
	\\
	\\
	
\multicolumn{7}{c}{E1: [buyticket,\_,\_,\{@AIM:BUY\_Success\}] $\leadsto$ E'1: [buyticket,\_,\_,\_] }	
\end{tabular}
\caption{\label{fig:auto:mut}Mutation of the automaton for Property~$1$ (left) and Property~$3$ (right)}
\end{figure}

\begin{example}[Automaton mutation]
Figure~\ref{fig:auto:mut} displays the application of a mutation on the automaton 
associated to Properties~$(1)$ and~$(3)$. We see that the mutated automaton makes it possible
to match test cases that would perform an attempt to purchase a ticket, before 
successfully logging in. 
\end{example}

\begin{definition}[Robustness coverage] A test suite is said to satisfy the robustness
coverage criterion for a property $P$ if and only if the mutated transition of each 
mutated automaton of property $P$ is covered by at least one test case of the test suite. 
\end{definition}

\begin{example}[Robustness coverage]
In order to activate the mutated event of Property~1, and thus, check the robustness
of the system w.r.t. it, the validation engineer can design the following test case:

\begin{center}
\tt \small
\begin{tabular}{|c|l|l|}
\hline
\bf Step & \bf Operation & \bf Expected behavior \\
\hline
\hline
1 & sut.buyTicket(TITLE1) &  @AIM:BUY\_Error\_Login\_First \\
\hline
2 & sut.login(REGISTERED\_USER,REGISTERED\_PWD) & @AIM:LOG\_Success \\
\hline
\end{tabular}
\end{center}

On a correct implementation, the system should not allow the first operation 
({\tt buyTicket}) to be performed successfully. If the implementation conforms to 
the model, then it is expected to activate an erroneous behavior of this operation
(as predicted by the model). If the implementation is incorrect, the {\tt buyTicket}
operation will return successfully and thus display a behavior that differs from 
the model. \\

In order to activate the mutated event of Property~3, the validation engineer can 
design the following test case:
\begin{center}
\tt \small
\begin{tabular}{|c|l|l|}
\hline
\bf Step & \bf Operation & \bf Expected behavior \\
\hline
\hline
1 & sut.login(REGISTERED\_USER,REGISTERED\_PWD) & @AIM:LOG\_Success \\
2 & sut.logout() & @AIM:LOG\_Logout \\
3 & sut.buyTicket(TITLE1) &  @AIM:BUY\_Error\_Login\_First \\
\hline
\end{tabular}
\end{center}
 
Similarly, on a correct implementation, the last operation ({\tt buyTicket}) should not 
succeed (as on the model). An incorrect implementation would allow this operation 
be performed successfully. 
\end{example}

We now present the experimental assessment of the approach proposed in this paper.

\vspace{-0.3cm}
\section{Experimentation}
\label{sec:experiments}

Our approach has been implemented into a specific framework~\cite{dcl+13:ip}, and 
applied in an industrial context during two national projects. This framework allows the
user to write properties, measure the coverage of a property with an existing test suite, 
and generate test scenarios to satisfy a given (selected) coverage 
criterion\footnote{see \url{http://vimeo.com/53210102} for a video demo of the tool}. 
We report here the usage of the tool and its evaluation by test engineers.
In addition, we present an experimental evaluation of the capabilities of our 
tool in terms of fault detection, especially focusing on the robustness testing 
criteria. \\

\noindent \textbf{Experimentations during industrial projects.}
Our approach has been initially designed during the ANR TASCCC 
project\footnote{\url{http://lifc.univ-fcomte.fr/TASCCC}}, and implemented into 
a prototype of the same name. This project was done in collaboration with (among others) 
the Smartesting company (technology provider with the CertifyIt test generator), and 
Gemalto (case study provider) and focused on the validation of smart card products 
for Common Criteria evaluations. In this context, the properties we designed aimed at
expressing functional security properties on a smart card case study. 
It has also been exploited during the ANR OSeP project\footnote{\url{http://osep.univ-fcomte.fr}}, 
also in partnership with 
the Smartesting company, and funded by the Armaments Procurements 
Agency. 
This project focused 
on online and offline techniques for testing security (cryptographic) components.
In both projects, we had the opportunity to address the following three questions:
{\it How easy it is to learn and use the test property language?~$(Q.~1)$} 
{\it How does this approach compare to a functional approach such as
Smartesting CertifyIt?~$(Q.~2)$}
{\it What is the interest of our industrial partners in our property-based 
testing approach?~$(Q.~3)$}

For our experiments, we considered the case studies developed by our partners. 
We started by designing test properties for the three considered case studies.
The language turned out to be easy to learn, as the number of combinations is
relatively small. With a little help from us, the validation engineers were able
to express some security properties in natural language, rephrase them in 
order to fit the allowed combinations of scopes/patterns, and choose the appropriate 
entities to express these properties~$[Q.~1]$. However, we noticed some pitfalls in its use. 
Indeed, the validation engineers tend to express test cases with the properties, 
e.g. by specifying sequences of events without taking care of the applicability 
of the sequence in general. 

For each case study, our industrial partners generated functional test cases using the 
Smartesting CertifyIt 
test generator. Thus, we measured the coverage of the test properties that were 
designed. For each of the three considered case studies, we noticed that some 
properties where not covered at all (no $\alpha-$transition of their associated 
automaton were activated), meaning that the test cases do not activate the scope
of the property~$[Q.~2]$. Such a process provides an interesting feedback on the coherence
of the model w.r.t. the properties and the completeness of the test suite. When 
a given coverage criterion is not satisfied, the validation engineer has the possibility 
to see which parts of the automaton were not covered, and focus his effort
on them~$[Q.~3]$.
In some cases, we detected some violations of the property during the test cases
executions. In this case, the coverage report shows that transitions leading to the 
error state of the property automaton is reached. This indicates that the model 
does not satisfy the property. There are two reasons for that, either the property
is too restrictive (e.g. in the cases described above), or the model 
is incorrectly designed (this was never the case). It is then mandatory to correct 
one or the other of these 
artefacts. In this end, this process can be (indirectly) used to validate the model, 
by checking that it behaves as described in the properties.  

In addition, we asked our industrial partners to evaluate the relevance of the 
produced tests (and, by extension, the relevance of the proposed coverage criteria). 
They pointed out the interest of the graphical visualization of the property automaton
which shows, and addresses, various configurations that one 
might forget when designing tests manually. \\

\noindent \textbf{Evaluation of fault detection capabilities.}
For our experiments, we have considered the eCinema model, along with additional 
properties that aim at illustrating various combinations of scopes and patterns. 
In addition to properties $(1)$-$(3)$ presented before, we considered three more specifying
that: ``before a successful ticket deletion, there eventually exists at least one successful 
ticket buying'' $(4)$, ``globally, a successful login precedes a successful logout'' $(5)$, 
and ``after having deleted all the tickets, it is not possible to delete one ticket unless buying 
one in the meantime'' $(6)$. 


We started by generating a test suite with the Smartesting CertifyIt test generator, and
completed this test suite with 11 tests in order to satisfy the different coverage 
criteria for each properties. We do not detail this process, as it is not the purpose of this paper; 
however, the interested reader may refer to~\cite{dcl+13:ip}. We then mutated the original 
eCinema model, by blindly using the following mutation operators, dedicated to OCL code: 
Simple Set Operator Replacement (SSOR) replacing a set operator by another,
Simple expression Negation Operator (SNO) negating an atomic condition
inside a predicate, Stuck-At-False (SAF) replacing predicates 
by false, and Action Deletion (AD) deleting a predicate of the postcondition (e.g.
to remove a variable assignment). 

We then run our test suite on these mutant models. We compared, at each step of each test, 
the expected result of each operation (given by the test case) with the actual returned values 
(this is how conformance is established on a concrete implementation), giving us a conformance 
verdict. In addition, we monitored the execution of 
each test on the automata associated to the considered test properties, to check if the 
observation was too weak to detect an actual property violation. Finally, we compared our test
suite with the original test suite computed by CertifyIt. 
Figure~\ref{fig:results_robustness} show the results we obtained. For each considered mutation 
operator we provide the overall number of mutants that were detected as: conform and not reaching 
an error state on the automaton (C-NE -- it is either an equivalent mutant or a mutant that violates the
property but it was not possible to observe it), non-conform but not reaching an error state 
(NC-NA -- it is the cases of mutants that are not related to the property or, at least, not killed 
for that reason), 
non-conform and reaching an error state (NC-E -- actual violations of the property detected 
using basic observations), conform but reaching an error state (C-A -- actual violations of the
property that have not been detected using basic observations, but could be detected by
more intrusive means such as monitoring). 

\begin{figure}[tb]
\footnotesize
\centering
\begin{tabular}{|c||c|c|c|c||c|c|c|c|}
\hline
Test suites &  \multicolumn{4}{c||}{Property-Based Testing} &  \multicolumn{4}{c|}{Smartesting CertifyIt} \\
\hline
Mutations / Verdicts & C-NE & NC-NA & NC-E & C-A & C-NE & NC-NA & NC-E & C-A \\
\hline
\hline
SSOR & 2 & 1 & 1 & 2 & 4 & 1 & 1 &  \\
\hline
SNO &  & 28 & 2 &  &  & 28 & & 2 \\
\hline
SAF &  & 31 & 1 &  & & 31 & 1 &   \\
\hline
AD & 6 & 12 & & 4  & 15 & 8 & & \\
\hline
\end{tabular}
\caption{\label{fig:results_robustness}Results of the mutant detection for the two considered test suites}
\end{figure}

Experimental results show that our technique is able to: $(1)$~build test cases that 
consist in operations leading to a violation of the property (see lines SSOR and AD
in which respectively 2 and 4 tests detect the mutant using 
the property automaton), $(2)$~build test cases that make property violations 
observable (see line SNO), and $(3)$~build new test cases that are likely to reveal 
other non-conformances (even if they are not related to the property), improving the 
overall efficiency of the test suite (see line AD). 


\vspace{-0.3cm}
\section{Related Work}
\label{sec:relatedworks}

The notion of property-based testing is often employed in the test generation 
context. Several approaches~\cite{GH99,TSL04,ADX01} deal with 
LTL formulae, that are negated and then given to a model-checker 
that produces traces leading to a counter-example of this property, 
and thus defining the test sequences. Our work improves these
approaches by defining both nominal and robustness test cases, 
aiming either at illustrating the property or checking the system's
robustness w.r.t. it. 
A recent work~\cite{fraser} defines the notion of property relevant test cases, 
introducing new coverage criteria that can be used to determine 
positive and negative test cases. 
Nevertheless, our approach proposes several differences. First, we do
not rely on LTL, but on a dedicated language easier to manipulate than 
LTL by non-specialists. Second, the notion of property-relevance is 
defined at the LTL level, whereas we rely on the underlying automata.
Finally, the relevance notion acts as an overlay to classical coverage 
criteria, while we propose new ones. 

Based on Dwyer's work, jPost~\cite{JPost} uses a property expressed in
a trace logic for monitoring an implementation. 
Similarly, in~\cite{rusu05a} the authors introduce the 
notion of observers, as ioSTS, that decide the satisfaction of the 
property and guide the test generation within the STG tool. Our work
differs in the sense that the coverage criteria are not only used as 
monitors for passive testing, but they can also be employed for 
active testing.

A lot of scenario-based testing works focus on extracting scenarios
from UML diagrams, such as the SCENTOR approach~\cite{scentor} or 
SCENT~\cite{ryser99practical} using statecharts. The SOOFT approach~\cite{tsai_sbt}
proposes an object oriented framework for performing scenario-based 
testing. In~\cite{binder}, Binder proposes the notion of round-trip scenario
test that cover all event-response path of an UML sequence diagram. 
Nevertheless, the scenarios have to be completely described.
Our approach proposes to automatically generate the test scenarios from
higher level descriptions of the properties the validation engineer wants 
to test. 
In~\cite{1083284}, the authors propose an approach for the automated scenario 
generation from environment models for testing of real-time reactive systems. The 
behavior of the system is defined as a set of events. The process relies on an attributed 
event grammar (AEG) that specifies possible event traces. Even if the targeted
applications are different, the AEG can be seen as a generalization of regular
expressions. Our approach goes further as it uses a property description language
that is close to a natural language.

\vspace{-0.3cm}
\section{Conclusion and Future Works}
\label{sec:conclusion}
In this paper, we have presented a model-based testing process based on 
test properties. These latter are expressed in a dedicated formalism that captures the 
dynamics of the system. Each property is translated into an automaton, for which new 
coverage criteria have been introduced, in order to illustrate the property.  
In addition, we propose to refine the automaton so as to exhibit specific 
transitions that are closely related to error traces that are not accepted 
by the property. This technique makes it possible to introduce a notion 
of robustness testing to ensure that the property is correctly implemented. 
We have tool-supported this approach to apply it to UML/OCL models~\cite{dcl+13:ip}. 
The advantages of this approach are twofold. Mainly, it provides a means
to produce test cases that can be directly related to the property. Such a
traceability makes it a suitable approach for industrial purposes. In addition, 
the automata and their refinements can be used to measure the coverage
of corner cases of a property for an existing test suite. 
This approach has been evaluated in the context of industrial projects, 
which gave us a very positive feedback on the usefulness of the coverage 
criteria, exhibiting specific sequences of operations one may want to 
consider when testing. Finally, notice that the proposed coverage criteria are 
not specific to UML/OCL and could be adapted to any other notation that would 
use the same notions of scope and patterns with a different representation of events.

For the future, we first plan to extend our
property language to introduce local variables. Such an extension would
greatly improve the expressiveness of our property language. However, 
this extension would imply the definition of data coverage criteria dedicated 
to the coverage of the properties of these values. Second, we are also 
investigating a way to efficiently generate the tests satisfying the coverage
criteria we proposed. For now, a solution using test scenarios has been 
implemented. However, the combinatorial unfolding of the scenarios 
compromises the full automation of the test generation approach.

\vspace{-0.3cm}
\bibliographystyle{eptcs}
\bibliography{biblio}
\end{document}